\documentclass[aps, prc, superscriptaddress, twocolumn]{revtex4-1} 
\usepackage{graphicx}
\usepackage{float}
\usepackage{amsmath}
\usepackage{subfigure}
\usepackage{siunitx}
\usepackage{verbatim}
\usepackage{color}
\usepackage{ulem}
\usepackage{multirow}
\usepackage{threeparttable}

\begin{document}
\title{High-Precision Half-life Measurement for the Superallowed Fermi $\beta^+$ Emitter $^{22}$Mg}

\author{M.~R.~Dunlop}\email{boudream@uoguelph.ca}\affiliation{Department of Physics, University of Guelph, Guelph, Ontario N1G 2W1, Canada}
\author{C.~E.~Svensson}\affiliation{Department of Physics, University of Guelph, Guelph, Ontario N1G 2W1, Canada}
\author{G.~C.~Ball}\affiliation{TRIUMF, 4004 Wesbrook Mall, Vancouver, British Columbia V6T 2A3, Canada}
\author{J.~R.~Leslie}\affiliation{Department of Physics, Queen's University, Kingston, Ontario K7L 3N6, Canada}
\author{C.~Andreoiu}\affiliation{Department of Chemistry, Simon Fraser University, Burnaby, British Columbia V5A 1S6, Canada}
\author{N.~Bernier}\affiliation{Department of Physics, University of British Columbia, Vancouver, British Columbia V6T 2A3, Canada}
\author{H.~Bidaman}\affiliation{Department of Physics, University of Guelph, Guelph, Ontario N1G 2W1, Canada}
\author{V.~Bildstein}\affiliation{Department of Physics, University of Guelph, Guelph, Ontario N1G 2W1, Canada}
\author{M.~Bowry}\affiliation{TRIUMF, 4004 Wesbrook Mall, Vancouver, British Columbia V6T 2A3, Canada}
\author{C.~Burbadge}\affiliation{Department of Physics, University of Guelph, Guelph, Ontario N1G 2W1, Canada}
\author{R.~Caballero-Folch}\affiliation{TRIUMF, 4004 Wesbrook Mall, Vancouver, British Columbia V6T 2A3, Canada}
\author{A.~Diaz~Varela}\affiliation{Department of Physics, University of Guelph, Guelph, Ontario N1G 2W1, Canada}
\author{R.~Dunlop}\affiliation{Department of Physics, University of Guelph, Guelph, Ontario N1G 2W1, Canada}
\author{A.~B.~Garnsworthy}\affiliation{TRIUMF, 4004 Wesbrook Mall, Vancouver, British Columbia V6T 2A3, Canada}
\author{P.~E.~Garrett}\affiliation{Department of Physics, University of Guelph, Guelph, Ontario N1G 2W1, Canada}
\author{G.~Hackman}\affiliation{TRIUMF, 4004 Wesbrook Mall, Vancouver, British Columbia V6T 2A3, Canada}
\author{B.~Jigmeddorj}\affiliation{Department of Physics, University of Guelph, Guelph, Ontario N1G 2W1, Canada}
\author{K.~G.~Leach}\affiliation{Department of Physics, Colorado School of Mines, Golden, Colorado 80401, USA}
\author{A.~D.~MacLean}\affiliation{Department of Physics, University of Guelph, Guelph, Ontario N1G 2W1, Canada}
\author{B.~Olaizola}\affiliation{Department of Physics, University of Guelph, Guelph, Ontario N1G 2W1, Canada}
\author{J.~Measures}\affiliation{TRIUMF, 4004 Wesbrook Mall, Vancouver, British Columbia V6T 2A3, Canada}
\author{C.~Natzke}\affiliation{Department of Physics, Colorado School of Mines, Golden, Colorado 80401, USA}
\author{Y.~Saito}\affiliation{TRIUMF, 4004 Wesbrook Mall, Vancouver, British Columbia V6T 2A3, Canada}
\author{J.~K.~Smith}\affiliation{Department of Physics, Reed College, Portland, Oregon 97202, USA}
\author{J.~Turko}\affiliation{Department of Physics, University of Guelph, Guelph, Ontario N1G 2W1, Canada}
\author{T.~Zidar}\affiliation{Department of Physics, University of Guelph, Guelph, Ontario N1G 2W1, Canada}

\begin{abstract}
A high-precision half-life measurement for the superallowed Fermi $\beta^+$ emitter $^{22}$Mg was performed at the TRIUMF-ISAC facility using a 4$\pi$ proportional gas counter. The result of $T_{1/2} = 3.87400 \pm 0.00079$ s is a factor of 3 more precise than the previously adopted world average and resolves a discrepancy between the two previously published $^{22}$Mg half-life measurements.
\end{abstract}
\maketitle

\section{Introduction}

Precision measurements of the $ft$ values for superallowed Fermi $\beta$-decay transitions between $J^\pi = 0^+$, $T=1$ isobaric analogue states provide fundamental tests of the electroweak interaction described by the Standard Model~\cite{Ha15}.  These transitions, which in leading order depend only on the vector part of the weak interaction, provide a stringent test of the Conserved-Vector-Current (CVC) hypothesis and a direct measure of the weak vector coupling constant,  $G_V$. In combination with the Fermi coupling constant, $G_F$, which is obtained from muon decay measurements, the superallowed transitions also currently provide the most precise determination of $V_{ud}$, the most precisely determined element of the Cabibbo-Kobayashi-Maskawa (CKM) quark mixing matrix~\cite{Ha15}. Combined with $V_{us}$ and $V_{ub}$, the top-row elements of the CKM matrix provide the most precise test of CKM unitarity with the result $|V_{ud}|^2 + |V_{us}|^2 + |V_{ub}|^2 = 0.9996 \pm 0.0005$~\cite{PDG16}. In addition to precision tests of the Standard Model, the superallowed Fermi transitions also set stringent limits on possible physics scenarios beyond the Standard Model, such as the existence of weak scalar currents~\cite{Du16}.

Experimentally, the superallowed $\beta$ decay $ft$-values are determined via measurements of the half-life, $T_{1/2}$, branching ratio, $BR$, and transition energy, $Q_{EC}$, of the $0^+\rightarrow0^+$ analogue transition. In addition to the experimental quantities, several theoretical corrections must be applied in order to obtain nucleus independent $\mathcal{F}t$ values. These ``corrected"-$\mathcal{F}t$ values for the superallowed $\beta^+$ emitters are given by~\cite{Ha15}:

\begin{equation}
\mathcal{F}t = ft(1+\delta^\prime_R)(1+\delta_{NS}-\delta_C) = \frac{K}{2G_V^2(1+\Delta^V_R)},
\end{equation}
where $K/(\hbar c)^6 = (8120.2776 \pm 0.0009) \times 10^{-10}$ GeV$^{-4}$s is a constant, $\Delta^V_R = (2.361 \pm 0.038)\%$ is a transition-independent radiative correction, $\delta_R^\prime$ and $\delta_{NS}$ are transition-dependent radiative corrections, and $\delta_C$ is the isospin-symmetry-breaking correction.

In the most recent survey of the world superallowed data~\cite{Ha15}, the constancy of the 14 transitions with $\mathcal{F}t$ values measured to a precision of at least 0.4\% was used to confirm the CVC hypothesis at the level of $1.2\times10^{-4}$. This constancy, and the resulting world-average superallowed $\overline{\mathcal{F}t}$ value used to determine $G_V$ and $V_{ud}$, are, however, dependent on the theoretical approach used to calculate the $\delta_C$ corrections. In the past two decades, many different approaches have been considered~\cite{Or95, Sa96, To02, To08, Mi08,Mi09, Au09, Li09, Ca09, Sa09, To10, Gr10, Sa11, Sa12}, with significant  model dependence of the resulting $\delta_C$ values. Such variations in $\delta_C$ directly affect the Standard Model tests of the CVC hypothesis, the determination of $V_{ud}$, and the test of CKM unitarity.

In previous evaluations of the world superallowed data~\cite{Ha09, Ha05}, the model dependence of the $\mathcal{F}t$-values arising from different theoretical approaches to the $\delta_C$ corrections was taken into account by assigning a systematic uncertainty to the world average $\overline{\mathcal{F}t}$ value. In particular, two sets of $\delta_C$ corrections were considered. In both approaches, $\delta_C$ is broken into $\delta_C = \delta_{C1} + \delta_{C2}$, where $\delta_{C1}$ accounts for the difference in configuration mixing in the parent and daughter $0^+$ states, and $\delta_{C2}$ accounts for the imperfect overlap between the daughter and parent radial wavefunctions. The $\delta_{C1}$ correction is obtained from a Shell Model (SM) calculation with the isospin non-conserving components of the Hamiltonian constrained to reproduce measured coefficients of the isobaric multiplet mass equation (IMME). In the first approach, the $\delta_{C2}$ correction is calculated using radial wavefunctions derived from a Woods-Saxon (SM-WS) potential~\cite{To08} whereas in the second approach Hartree-Fock (SM-HF) eigenfunctions are used~\cite{Or85, Ha09, Or89, Or95}. The model-dependence of the $\delta_C$ correction was taken into account by averaging the $\overline{\mathcal{F}t}$ values from the SM-WS and SM-HF calculations, and assigning a systematic uncertainty based on the difference between the two $\overline{\mathcal{F}t}$ values.

In the most recent evaluation of the world superallowed data~\cite{Ha15}, however, only the SM-WS $\delta_{C2}$ correction was used in the $\mathcal{F}t$ calculation and no model dependent systematic uncertainty arising from the $\delta_C$ correction was assigned to the $\overline{\mathcal{F}t}$ value. This choice was motivated by the better agreement with the CVC hypothesis obtained with the SM-WS $\delta_{C2}$ corrections as well as a recent measurement of the $^{38}$Ca branching ratio~\cite{Pa14} which resulted in a better agreement between the experimentally determined ratio of $ft$ values of the ``mirror" superallowed transitions $^{38}$Ca $\rightarrow ^{38m}$K and $^{38m}$K $\rightarrow$ $^{38}$Ar with the calculated $ft$ ratio using the SM-WS approach~\cite{Ha15}. 

\begin{figure}
\includegraphics[scale=0.33]{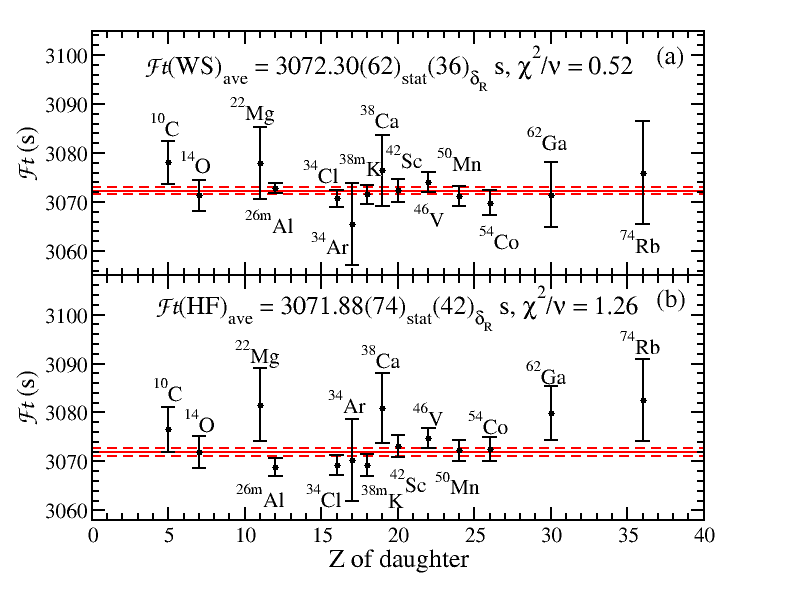}
\caption{(Color online) A plot of the 14 precisely measured superallowed $\mathcal{F}t$ values obtained in the most recent evaluation using (a) the SM-WS $\delta_C$ corrections of Ref.~\cite{Ha15} and (b) the SM-HF $\delta_C$ corrections of Ref.~\cite{Ha09}. In both panels, the solid line represents the weighted average and the dashed lines represent the $\pm 1\sigma$ uncertainties.}
\label{fig:Ft_WSHF}
\end{figure}

The 14 most precisely measured $\mathcal{F}t$ values from the most recent evaluation~\cite{Ha15} are plotted for both sets of $\delta_C$ corrections in Figure~\ref{fig:Ft_WSHF}.  Although the SM-WS corrections show better agreement with the CVC hypothesis, with a $\chi^2/\nu = 0.52$, the $\mathcal{F}t$ values obtained when the SM-HF $\delta_C$ corrections~\cite{Ha09} are used yield a $\chi^2/\nu = 1.26$ for 13 degrees of freedom, which has a probability of 23\% for statistically independent data. Furthermore, as can be seen in Figure~\ref{fig:Ft_WSHF}, the larger $\chi^2/\nu$ obtained when the SM-HF corrections are used originates primarily from four of the least precisely determined $\mathcal{F}t$ values included in the evaluation, namely $^{22}$Mg, $^{38}$Ca, $^{62}$Ga, and $^{74}$Rb. If only the 9 transitions with $\mathcal{F}t$ values determined to 0.15\% or better are retained, one obtains $\overline{\mathcal{F}t}_{WS} = 3072.20(63)_{stat}(36)_{\delta_R^\prime}$ with $\chi^2/\nu = 0.67$ and $\overline{\mathcal{F}t}_{HF} = 3071.43(76)_{stat}(42)_{\delta_R^\prime}$ with $\chi^2/\nu = 1.00$, both of which are consistent with the CVC hypothesis but with central values differing by 0.77~s, equivalent to the entire statistical uncertainty of the world superallowed dataset. For the high-$Z$ cases of $^{62}$Ga and $^{74}$Rb, the $\mathcal{F}t$ uncertainties are dominated by the theoretical corrections, but for the $T_z = -1$ emitters $^{22}$Mg, $^{34}$Ar, and $^{38}$Ca, the $\mathcal{F}t$ uncertainties are currently dominated by the precision of the experimental data. Improved measurements for these $T_z = -1$ emitters are thus crucial for testing the model dependence of the isospin-symmetry-breaking corrections in superallowed Fermi $\beta$ decays. 

For the case of $^{22}$Mg, the uncertainty in the $\mathcal{F}t$ value is dominated by the experimental uncertainties in the branching ratio and half-life measurements. The currently adopted half-life of $^{22}$Mg, $T_{1/2} = 3.8752 \pm 0.0024$ s~\cite{Ha15}, is determined from two measurements, $T_{1/2} = 3.8755 \pm 0.0012$~s~\cite{Ha03} and $T_{1/2} = 3.857 \pm 0.009$~s~\cite{Ha75}. The disagreement between these two measurements, with a $\chi^2/\nu = 4.0$, leads to an inflation in the uncertainty of the adopted world-average half-life for $^{22}$Mg by a factor of 2~\cite{Ha15}. In this paper, we report a new measurement of the $^{22}$Mg half-life with a precision of 0.02\%. This new measurement is in agreement with, but 1.5 times more precise than, the measurement reported in Ref.~\cite{Ha03}. The agreement of the two high precision measurements excludes the older, less precise, result of Ref.~\cite{Ha75} and results in an improvement in the precision of the world-average half-life by more than a factor of 3.

\section{Experiment}

The experiment was performed at TRIUMF's Isotope Separator and Accelerator (ISAC) facility where the Isotope Separation On-Line (ISOL) technique is used to produce radioactive ion beams (RIBs)~\cite{TRIUMF}. A 40 $\mu$A beam of 480 MeV protons from TRIUMF's main cyclotron impinged on a SiC target to produce spallation products. The target was coupled to the Ion Guide Laser Ion Source (IGLIS) to produce intense beams of laser-ionized $^{22}$Mg, while suppressing surface ionized contaminants, such as $^{22}$Na, by a factor of $10^5$-$10^6$~\cite{Ra14}. A high-resolution mass separator was then used to select a beam of singly ionized $A = 22$ products which included $^{22}$Mg at $\sim10^5$ ions/s and a remaining contaminant of $^{22}$Na at $\sim10^4$ ions/s which was delivered to the experimental hall as a 30 keV ion beam.

The half-life measurements were performed using a 4$\pi$ continuous-flow gas proportional counter which detects $\beta$ particles with near 100\% efficiency~\cite{Du16}. The beam was implanted under vacuum into a thick (17.2 $\mu$m) Al layer of an aluminized Mylar tape~\cite{La15} for a duration of $0.6$-$0.7$~s in order to build up a source of $^{22}$Mg, after which the beam was deflected after the mass separator. In order to avoid space charge effects, methane gas was continually flushed through the gas counter at a rate of approximately 0.5 cc/min and the maximum counting rate in the gas counter was limited to $\leq 14$ kHz by allowing the $^{22}$Mg sample to ``cool" for $\sim$ 2 s. The tape was then moved into the centre of the gas counter in $\sim 1$~s and the decay was measured for $90$-$100$ s, corresponding to approximately 25 half-lives of $^{22}$Mg. The amplified and discriminated pulses from the gas counter were fanned to two LeCroy 222N gate-and-delay generators where two fixed and nonextendible dead-times of approximately 3~$\mu$s and 4~$\mu$s were applied. The dead-time affected data were then multiscaled in two multichannel scaler (MCS) modules with a Stanford Research Systems model DS335 temperature stabilized precision clock used to provide the time standard. The two dead-times were interchanged between the two MCS modules throughout the experiment to investigate any potential systematic effects. The $^{22}$Mg decay data were binned into 250 channels with channel dwell times of either 0.40~s or 0.36~s. Following the decay, the tape was moved into a tape disposal box in order to remove any long-lived contaminants out of view of the detector. Following the experiment, the frequency of the time standard, set to a nominal value of 100 kHz, was measured to be 99.99978~kHz and was found to be stable to $\pm 0.1$~ppm over a period of 24 hours. 

\section{Analysis}

\begin{figure}
\begin{center}
\includegraphics[scale=0.3]{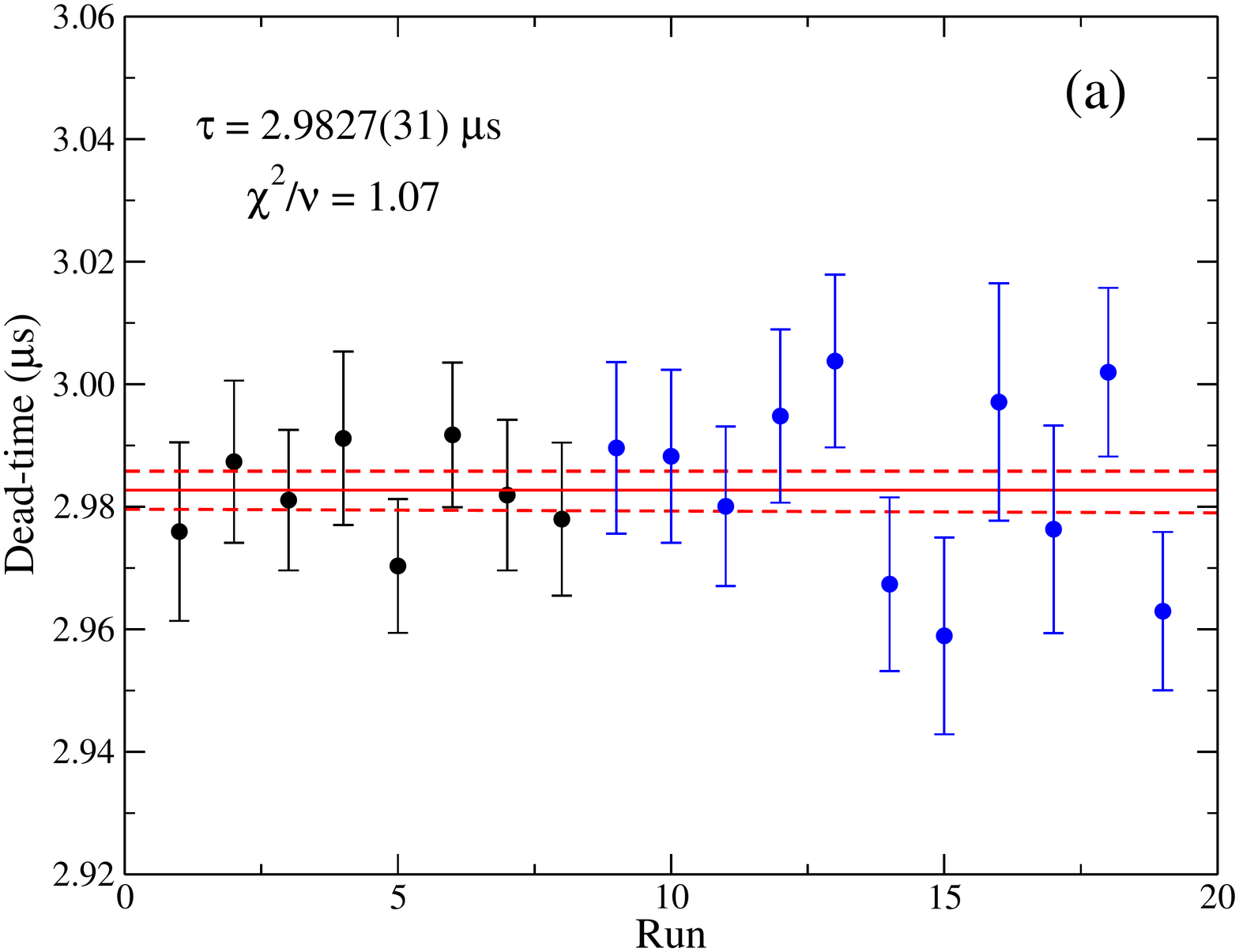}
\includegraphics[scale=0.3]{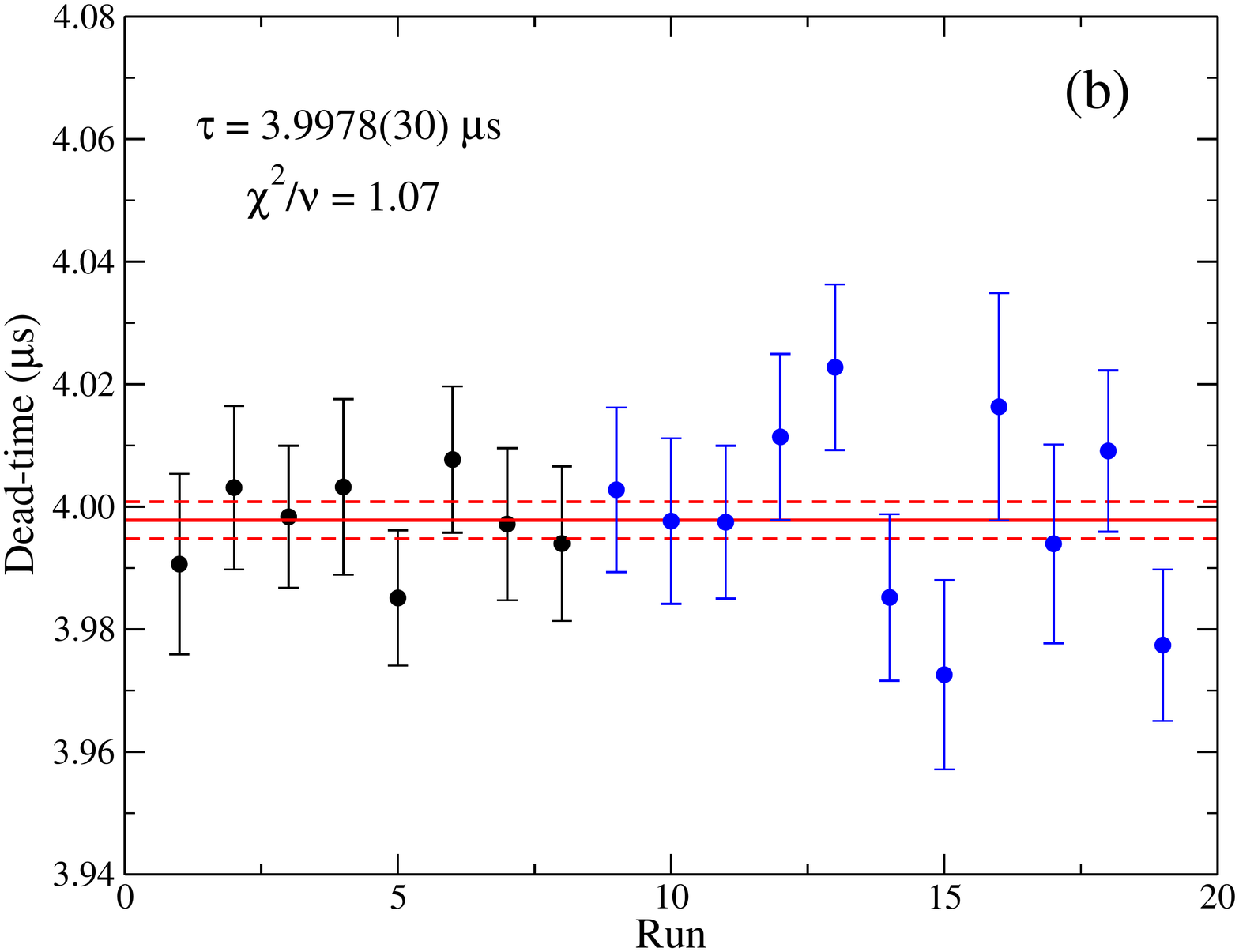}
\caption{(Color online) Measurements of the applied dead-time with nominal values of (a) 3 $\mu$s and (b) 4 $\mu$s. The weighted average of the data points and corresponding $\pm 1\sigma$ uncertainty is given by the solid and dashed lines, respectively.  Measurements performed before the $^{22}$Mg half-life experiment are given in black and the measurements performed after the experiment are given in blue.}
\label{fig:DT}
\end{center}
\end{figure}

Measurements of the applied dead-time were performed before and after the experiment using the source-plus-pulser technique~\cite{Ba65}. A plot of the dead-time measurements for each run is shown in Figure~\ref{fig:DT}. Following a small inflation of the statistical uncertainty by $\sqrt{\chi^2/\nu} = \sqrt{1.07}$ for both datasets, the two applied dead-times were determined to be 2.9827(32) $\mu$s and 3.9978(31) $\mu$s, respectively. 

\begin{figure}
\begin{center}
\includegraphics[scale=0.3]{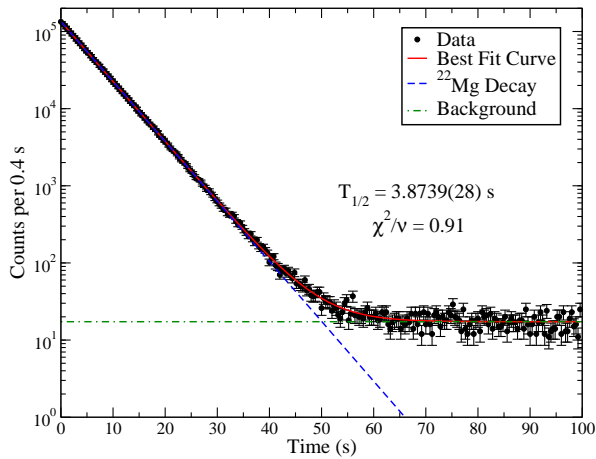}
\caption{(Color online) Dead-time corrected data points (black) and best fit curve (red) for a typical run. The fit to the data includes a $^{22}$Mg decaying component as well as a constant background component.}
\label{fig:DecayCurve}
\end{center}
\end{figure}

\begin{figure}
\begin{center}
\includegraphics[scale=0.3]{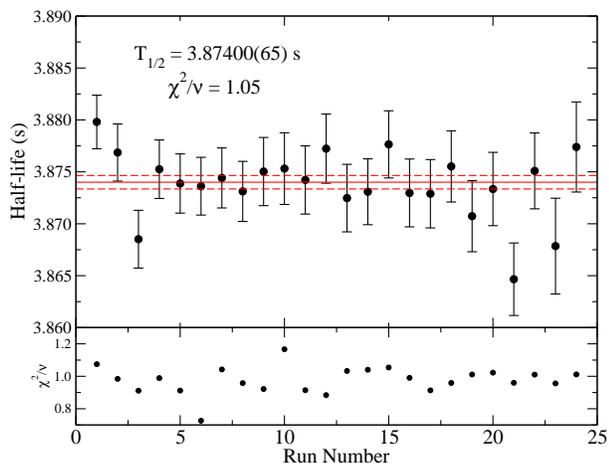}
\caption{(Color online) Run by run distribution of the half-life measurements. The corresponding $\chi^2/\nu$ of the fit for each run is given in the lower panel.}
\label{fig:T12Runs}
\end{center}
\end{figure}

Each cycle of the $^{22}$Mg decay data was analyzed individually in order to assess the quality of the data. Cycles with a considerable drop in the total number of counts, corresponding to a drop in the beam intensity due to interruptions of the proton beam, were removed. Cycles in which spurious noise signals in the gas counters occurred were also removed from the analysis. Following the inspection of the individual cycles, a total of 681 good cycles from 24 runs were used in the final analysis corresponding to 97\% of the total data acquired during the experiment. The data from the individual cycles were dead-time corrected and, for a given run, the dead-time corrected data from each cycle was summed. These data were then fit using a Poisson log-likelihood function in a Levenberg Marquardt $\chi^2$ minimization method~\cite{Gr05, Fi11}. The activity was fit to an exponentially decaying function which included the primary $^{22}$Mg component and a constant background component. The dead-time corrected data and best fit curve for a typical run is shown in Figure~\ref{fig:DecayCurve}. The deduced half-life from each of the MCS modules for a given run was then averaged. The deduced half-lives for each run are shown Figure~\ref{fig:T12Runs} and yield a weighted average half-life of $3.87400 \pm 0.00065$~s with a $\chi^2/\nu$ of 1.05 over the 24 runs.    

\subsection{Contaminants}

Although the IGLIS ion source suppresses the surface ionized $^{22}$Na contaminant in the beam by a factor of $10^5$-$10^6$, it is still delivered to the experimental station at a rate of approximately $10^4$ ions/s. Since $^{22}$Na has a half-life of $T_{1/2} = 2.6029(8)$~y~\cite{22Na}, the activity from $^{22}$Na is very well approximated by a constant over the $\sim 100$~s decay period and is thus accounted for in the constant background component of the fit function. In order to ensure that the $^{22}$Na activity has no effect on the deduced $^{22}$Mg half-life, the constant background component in the fit was replaced with an exponentially decaying function with the half-life of $^{22}$Na. Re-fitting the data with the decay of $^{22}$Na results in a change in the $^{22}$Mg half-life at the $10^{-9}$ s level and is thus entirely negligible at the level of precision reported here. Similarly, the ``grow-in" of $^{22}$Na activity as the daughter of $^{22}$Mg decay could, in principle, also affect the deduced half-life. The data was thus again re-fit with an additional term corresponding to the grow-in activity of the $^{22}$Na daughter. A change in the $^{22}$Mg half-life at the $10^{-10}$ s level was obtained and is also entirely negligible relative to the statistical uncertainty.

\begin{figure}[h]
\begin{center}
\includegraphics[scale=0.3]{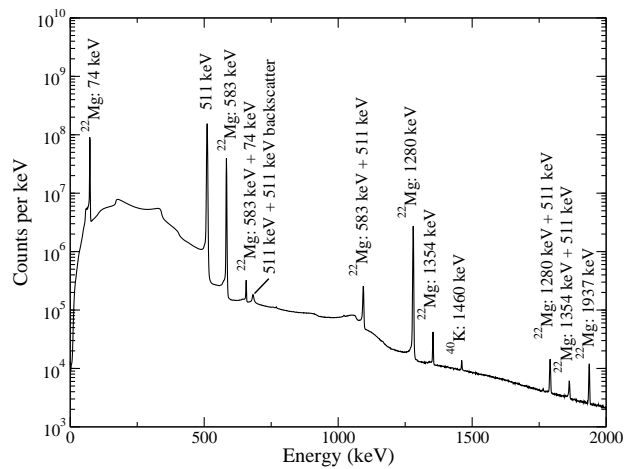}
\end{center}
\caption{$\gamma$-ray energy spectrum in $\beta$-$\gamma$ coincidence recorded with the GRIFFIN spectrometer. The absence of characteristic $\gamma$-rays from potential isobaric contaminants was used to set limits on contributions to the activity in the gas counter.}
\label{fig:Energy}
\end{figure}

Contributions from several other potential in-beam isobaric contaminants that would not be resolved by the mass separator were also considered, although with the combination of much lower ionization efficiency for these contaminants as well as the surface ion suppression provided by IGLIS, none were expected to be present in the beam. Potential contaminants included: $^{22}$O, $^{22}$F, $^{21}$Na and $^{21}$F (which could, in principle, be delivered at $A=22$ as molecular beams with hydrogen), and $^{44}$K$^{2+}$ (whose charge to mass ratio is 22). During the experimental running time, the $A=22$ RIB was also delivered to the GRIFFIN $\gamma$-ray spectrometer~\cite{Sv14, Ga17} which is located next to the gas counter. A $\beta$-$\gamma$ coincidence $\gamma$-ray energy spectrum, corresponding to a subset of the data taken during the experiment, is shown in Figure~\ref{fig:Energy}. The absence of any of the characteristic photopeaks from the decay of contaminants suggests that no additional isobaric contamination was present in the beam. Nonetheless, the $\gamma$-ray data was used to set upper limits on the contribution from each contaminant by fitting photopeaks of fixed full-width half maximum (FWHM) at the expected location of the characteristic $\gamma$-rays from each contaminant. An upper limit on the activity of the contaminant determined in this way was then included in the half-life fitting procedure, with its half-life and initial activity of the contaminant included as fixed parameters in the fit. For the unphysical cases in which the central value of the fitted peak areas was negative, the Gaussian probability distribution was integrated over the physical region of positive counts and the $1 \sigma$ upper limit was deduced by determining the number of counts corresponding to 68\% of the area of the positive count region of the probability distribution. The deduced upper limits, as well as the change in the $^{22}$Mg half-life when the individual contaminants were included in the fitting procedure, are shown in Table~\ref{table:contams}. The resulting changes in the $^{22}$Mg half-life from the inclusion of each contaminant were added in quadrature yielding a total contribution of 0.00011~s which we assign as a systematic uncertainty.

\begin{table}[h]
\begin{center}
\caption{Investigation of potential contaminants in the $A=22$ radioactive ion beam. The upper limit on the contaminant corresponds to its activity relative to $^{22}$Mg at the beginning of the counting period. The change in the $^{22}$Mg half-life is relative to the deduced $^{22}$Mg half-life when no contaminants are included in the fitting function.}
\vspace{14 pt}
\label{table:contams}
\setlength{\tabcolsep}{0.2em}
\begin{tabular}{c  c   c   c }
\hline \hline
\multirow{2}{*}{} Potential & Half-life & Upper Limit & Change in the   \\
			 Contaminant & (s) & on the Activity & $^{22}$Mg Half-life  \\ 
 			& \cite{NNDC, Fi17} & Relative to $^{22}$Mg & (s)   \\ \hline
$^{22}$O & 2.250(90)  &$1.1 \times 10^{-4}$ & -0.00010   \\ 
$^{22}$F & 4.230(40) & $1.4 \times 10^{-6}$& $5.6 \times 10^{-7}$ \\ 
$^{21}$Na & 22.448(8) & $1.5 \times 10^{-6}$  & 0.00004 \\ 
$^{21}$F & 4.158(20) & $9.9 \times 10^{-6}$ & 0.00001 \\ 
$^{44}$K & 1327.8(114) &$1.7 \times 10^{-8}$ & $4.2 \times 10^{-8}$ \\ 
\hline
 & & Total & 0.00011 \\
\hline \hline
\end{tabular}
\end{center}
\end{table}


\subsection{Rate Dependent Effects}

\begin{figure}[h]
\begin{center}
\includegraphics[scale=0.3]{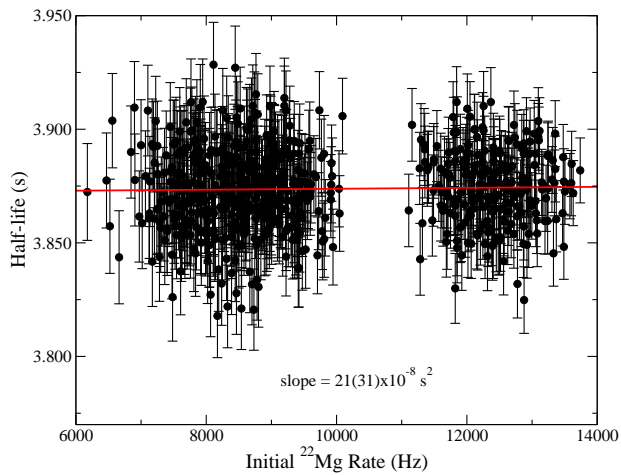}
\caption{(Color online) The half-life of $^{22}$Mg as a function of the initial count rate in the gas counter. The slope as determined from a weighted linear regression is shown in red and is consistent with zero.}
\label{fig:InitRate}
\end{center}
\end{figure}

During the experiment the rate in the gas counter was limited to $\leq$ 14 kHz in order to avoid space charge effects which can affect the half-life measurement. Nonetheless, we investigated a possible dependence of the half-life on the count rate in the gas counter. A plot of the half-life deduced from each cycle as a function of the initial rate is shown in Figure~\ref{fig:InitRate}. A weighted linear regression yields slope of $21(31)\times 10^{-8}$~s$^{2}$. This is consistent with zero, indicating no effect on the deduced half-life over the range of rates in the gas counter used in this experiment. 

\begin{figure}
\begin{center}
\includegraphics[scale=0.3]{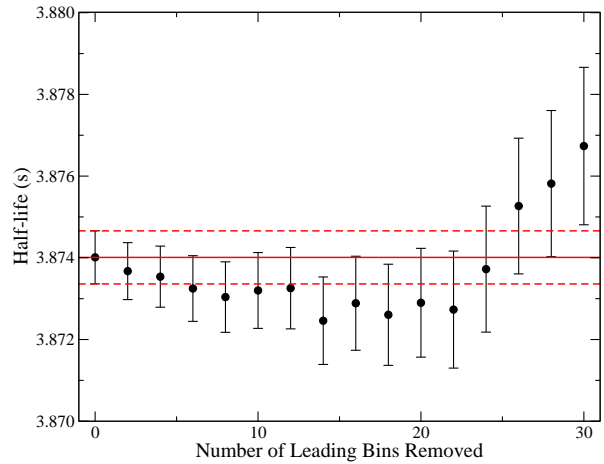}
\caption{(Color online) The half-life of $^{22}$Mg with the removal of leading channels for the data. Since each data point contains all of the data to the right of it, the data points are strongly correlated and are not expected to be scattered about the mean. The removal of 30 channels corresponds to approximately 3 half-lives of $^{22}$Mg, or $\sim 87.5\%$ of the data. No trend indicative of rate dependent effects is observed.}
\label{fig:Chop}
\end{center}
\end{figure}

Additional investigation of possible rate dependent effects was performed by systematically removing leading channels from the decay curve and re-fitting the remaining data. The deduced half-life as a function of the number of leading channels removed is shown in Figure~\ref{fig:Chop}, with no statistically significant change in the half-life observed.

\subsection{Systematic Uncertainties}

\begin{figure}[h]
\begin{center}
\includegraphics[scale=0.3]{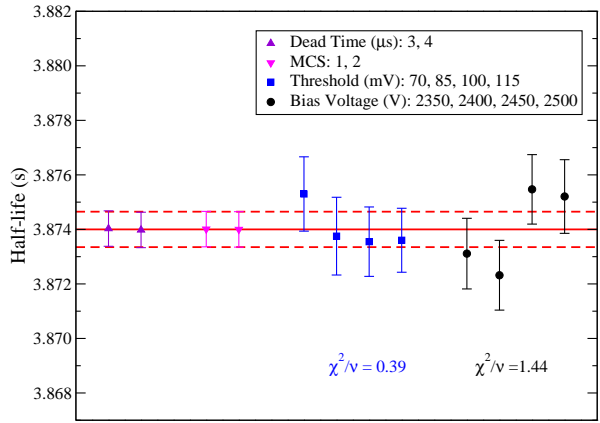}
\caption{(Color online) Half-life measurements of $^{22}$Mg grouped according to the different experimental running conditions. No $\chi^2/\nu$ is calculated for the dead-time and MCS groupings as these data are not statistically independent. The solid red line represents the deduced $^{22}$Mg half-life while the red dashed lines represent the corresponding $\pm 1\sigma$ uncertainties. Since the $\chi^2/\nu$ is greater than 1 for the different bias voltages used, the statistical uncertainty is inflated by $\sqrt{1.44}$ and assigned as a systematic uncertainty.}
\label{fig:Sys}
\end{center}
\end{figure}

Since the data used in the analysis are the averages of the data from the two MCS modules, with two different imposed dead-times, both the data from the individual MCS modules and the data with common dead-times were compared. As shown in Figure~\ref{fig:Sys}, both of the MCS modules and the two different dead-time values yield half-life results in complete agreement.

Following the completion of each run, the experimental running conditions were varied in order to investigate potential systematic effects arising from the choice of running conditions. The grouping of the data according to the different running conditions, which included different applied bias voltage (2350 V, 2400 V, 2450 V, and 2500 V) on the gas counter and different threshold voltages (70 mV, 85 mV, 100 mV, and 115 mV) are also shown in Figure~\ref{fig:Sys}. While the grouping by threshold settings yields $\chi^2/\nu = 0.39$, the grouping by bias voltage gives $\chi^2/\nu = 1.44$. Although a $\chi^2/\nu = 1.44$ for three degrees of freedom is expected 23\% of the time for statistically independent data, we follow the conservative approach recommended by the Particle Data Group~\cite{PDG16} and inflate the statistical uncertainty by the largest $\chi^2/\nu$ value. For this analysis, we thus inflate the uncertainty by $\sqrt{1.44}$ and assign 0.00043~s as a systematic uncertainty. 

Finally, the dead-times were also varied within $\pm 1 \sigma$ of their measured values and the data re-fit in order to investigate any variations in the half-life. The resulting change in the half-life of 0.00004~s was included as an additional systematic uncertainty.



The final $^{22}$Mg half-life result from this work is thus: 

\begin{equation}
\begin{split}
T_{1/2} &= 3.87400 \pm 0.00065_{stat} \pm 0. 00043_{sys}\\
& \pm 0.00011_{contam} \pm 0.00004_{DT}~\mathrm{s}\\
&= 3.87400 \pm 0.00079~\mathrm{s.}
\end{split}
\label{eqn:FinalT12}
\end{equation}

\section{Conclusions}

\begin{figure}[h]
\vspace{2mm}
\begin{center}
\includegraphics[scale=0.3]{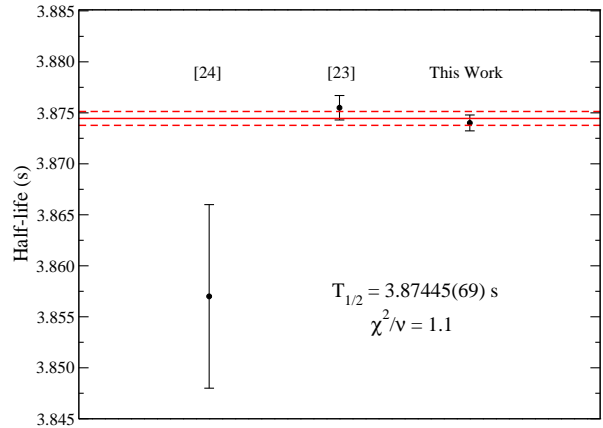}
\caption{(Color online) Half-life measurements for $^{22}$Mg from Refs.~\cite{Ha75}, ~\cite{Ha03}, and the current work. The red solid line represents the weighted average of the two precise measurements and the dashed red lines represent the corresponding $\pm 1\sigma$ uncertainty. The result of Ref.~\cite{Ha75} is more than a factor of 10 less precise than the result presented in the current work and is thus excluded from the weighted average~\cite{Ha15}}
\label{fig:UpdatedT12}
\end{center}
\end{figure}

A high-precision half-life measurement for the superallowed $\beta^+$ emitter $^{22}$Mg was performed yielding a half-life of 3.87400 $\pm 0.00079$~s, which is now the most precise measurement of the $^{22}$Mg half-life. As this result is more than a factor of 10 more precise than the value of $T_{1/2} = 3.857 \pm 0.009$~s reported in Ref.~\cite{Ha75}, we follow the procedure of Ref.~\cite{Ha15} and exclude the result of Ref.~\cite{Ha75} in the final averaging, obtaining a new world-average of $T_{1/2} = 3.87445 \pm 0.00069$ s with a $\chi^2/\nu$ of 1.1 from a weighted average of the current measurement and that of Ref.~\cite{Ha03}, as shown in Figure~\ref{fig:UpdatedT12}. This represents an improvement in the precision of the world average half-life by more than a factor of 3 and resolves the discrepancy between the two previously published half-life measurements. Including this new half-life measurement, as well as a recent measurement of $4781.40\pm 0.22$ keV~\cite{Le17} for the $Q_{EC}$ value between the $^{22}$Mg and $^{22}$Na ground states, with the previously evaluated superallowed data compiled in Ref.~\cite{Ha15} yields an updated $ft$ value of $3051.0 \pm 6.9$~s for $^{22}$Mg superallowed decay. The uncertainty of the $^{22}$Mg $ft$ value is now completely dominated by the uncertainty of the superallowed branching ratio. An improvement in the precision of this branching ratio will thus be critical to compare the tests of the CVC hypothesis using the SM-HF and the SM-WS $\delta_{C}$ corrections for~$^{22}$Mg.

\begin{acknowledgements}
This research was supported by the Natural Sciences and Engineering Research Council of Canada (NSERC) and the Canada Research Chairs Program. The GRIFFIN spectrometer used for the $\gamma$-ray analysis was jointly funded by the Canada Foundation for Innovation (CFI), TRIUMF, and the University of Guelph. TRIUMF receives federal funding via a contribution agreement through the National Research Council of Canada.
\end{acknowledgements}

\end{document}